\documentstyle[buckow1,12pt,epsfig]{article}
\begin {document}
\def\titleline{
\vspace{-18mm}
\rightline{\small HUB-EP-98/05, KANAZAWA 98-01}
\vspace{-10pt}
\rightline{\small January 14, 1998}
Hot Electroweak Matter
\newtitleline 
    Near to the Critical Higgs Mass\footnote{Talk 
    at Buckow Symp. 97
    given by E.-M.~Ilgenfritz, 
    supported by DFG under grant Mu932/1-4}
 }
\def\authors{
M.~G\"urtler\1ad, E.-M.~Ilgenfritz{\2ad  \3ad}, A.~Schiller\1ad, C.~Strecha\1ad}
\def\addresses{
\1ad
Institut f\"ur Theoretische Physik, Universit\"at Leipzig, 
D-04109 Leipzig, Germany \\
\2ad
Institut f\"ur Physik, Humboldt-Universit\"at,  
  D-10115 Berlin, Germany \\
\3ad
Institute for Theoretical Physics, Kanazawa University, 
Kanazawa 920, Japan 
}
\def\abstracttext{
We discuss the end of the 
first order phase transition
and the bound state spectrum, both
at weak transition and at the crossover. 
}
\makefront
\vspace{-14mm}
\section{Introduction}
\vspace{-3mm}
Two years ago~\cite{buckow95},
we have presented here first results of our group
obtained within the $3D$ approach,
comparing the properties of the thermal phase transition in the 
$SU(2)$ Higgs model at small coupling (Higgs mass) $M_H^*=35$ GeV
with a more realistic $M_H^*=70$ GeV.\footnote{$M_H^*$, together with
the $3D$ gauge coupling, is labelling our
lattice data. For relation to the zero-temperature 
Higgs mass, see Ref.~\cite{wirNP97}.} 
The endpoint of the phase transition
and the physics near to the corresponding
temperature $T_c(m_H^{crit})$ is the topic of this talk. 

In the meantime, 
due to efforts of three groups~\cite{kajantie,bi,wirNP97} 
using the $3D$ approach, 
various aspects of the high temperature electroweak 
phase transition (as latent heat and interface tension) have been
explored in this model within this
span of Higgs masses where the character of the transition changes 
drastically.
The accuracy is  
inaccessible to $4D$ Monte Carlo simulations~\cite{4D}, although the results
are consistent with each other 
where they can be compared~\cite{rummu,interface}. 

The interest in the properties of this phase
transition resulted from the hope to work out 
- within the standard model -
a viable mechanism for the generation of baryon asymmetry of the 
universe at the electroweak scale (see Z.~Fodor~\cite{fodor}).
It turned out, however, that in the case of the standard model,  
taking the lower bound of the Higgs mass into account,
it is at most weakly first order.
Together
with the small amount of CP violation in the standard model this has ruled
out the most economic scenario of BAU generation. 

Using dimensional reduction \cite{generic} the
simplest effective $3D$ $SU(2)$ Higgs theory has the action
\begin{equation}
S_3 = \int d^3 x \Big(\frac{1}{4} F_{\alpha \beta}^b F_{\alpha
    \beta}^b + (D_{\alpha} \phi)^+ (D_{\alpha} \phi)
     + m_3^2 \phi^+
  \phi + \lambda_3 (\phi^+ \phi)^2 \Big)
\label{eq:S3Dcontinuum}
\end{equation}
with dimensionful, renormalisation group invariant 
couplings $g_3^2$ and $\lambda_3$ and a running mass squared $m_3^3(\mu_3)$.
It is
related to the corresponding lattice model with the action
\begin{equation}
S = \beta_G \sum_p \big(1 - \frac{1}{2} \mbox{tr} U_p \big) -  
 \beta_H \sum_{x,\alpha}
       \frac{1}{2} \mbox{tr} (\Phi_x^+ U_{x, \alpha} \Phi_{x + \alpha}) 
 + \sum_x  \big( \rho_x^2 + \beta_R (\rho_x^2-1)^2 \big)
\label{eq:S3Dlattice}
\end{equation}
and
$\rho_x^2=\frac12\mbox{tr}(\Phi_x^+\Phi_x)$ 
through the following relations (clarifying the meaning of $M_H^*$)
\begin{eqnarray}
\beta_G = \frac{4}{a g_3^2}\,, \ \  
\beta_R = \frac{\lambda_3}{g_3^2} 
\frac{\beta_H^2}{\beta_G}
= \frac18 \left( \frac{M_H^*}{80\; \mbox{GeV}}\right)^2 
\frac{\beta_H^2}{\beta_G}\,, \ \  
\beta_H=\frac{2 (1-2\beta_R)}{6+a^2 m_3^2} \, .
\label{eq:couplings}
\end{eqnarray}
The coupling 
parameters in the lattice action
can be expressed in terms of $4D$ couplings and masses.
The bare mass squared is related to the
renormalised $m_3^2(\mu_3)$ (we choose $\mu_3=g_3^2$) through a 
lattice two-loop calculation~\cite{laine}.
Most of the $3D$ numerical investigations 
have been done for the $SU(2)$ Higgs model. 
The results obtained in the $3D$ approach indicate the validity of the
dimensional reduction near to 
the transition temperature for Higgs masses between 
$30$ and $100$ GeV.

Our simulations have been done on
the DFG-sponsored Quadrics QH2 in Bielefeld, 
on a CRAY-T90 and a
Quadrics Q4 at HLRZ in J\"ulich. 
Algorithmic details are described in Ref.~\cite{wirNP97}.
In the course of time, we have
explored the parameter range
$M_H^*= 35,57,70,74,76,80$ GeV, looking for the strength of the
transition, 
checking the approach to the continuum limit
by varying
$\beta_G =8, 12, 16 \propto 1/a$
and comparing with perturbation theory.
The simulations at the last three 
parameter values $M_H^*$ have been devoted exclusively to 
locate
the endpoint of the transition line \cite{endpoint} in the $m_H - T$ plane.  
The lattice size usually varied from $20^3$ to $80^3$.
In order to unambiguously locate the critical Higgs
mass we had to increase the lattice to $96^3$ 
simulating at 
$M_H^*=74$ GeV, the nearest to the critical value.

It has been essential to make extensive use of the 
Ferrenberg-Swendsen multihistogram technique. 
Data obtained at various $M_H^*$ and $\beta_H$ 
have been subject to global analysis by reweighting to
other values, eventually extending  
the hopping parameter to
complex values.
The method constructs 
an optimal estimator of the spectral density
of states (at fixed $L$ and $\beta_G$) occurring in the 
representation of 
the partition function
\begin{equation}
Z(L,\beta_G,\beta_H)=
 \int dS_1 dS_2 D_L(S_1,S_2,\beta_G)
  \exp\left(L^3 (\beta_H S_1 -\beta_H^2 S_2)\right)
\label{eq:sos_from_dos}
\end{equation}
from the {\it measured} double-histograms in the two variables
\begin{equation}
  S_1=3  \sum_{x,\alpha}
         \frac{1}{2} \mbox{tr} (\Phi_x^+ U_{x, \alpha} \Phi_{x + \alpha})
         \,\, , \, \, S_2= \frac{\lambda_3}
  {g_3^2 \beta_G}(\sum_x \rho_x^4-2 \sum_x \rho_x^2) \, .
\label{eq:explain}
\end{equation}
\section{Finding $m_H^{crit}$}
\begin{figure}[!htb]
  \begin{minipage}[t]{7.3cm}
    \begin{center}
      \epsfig{file=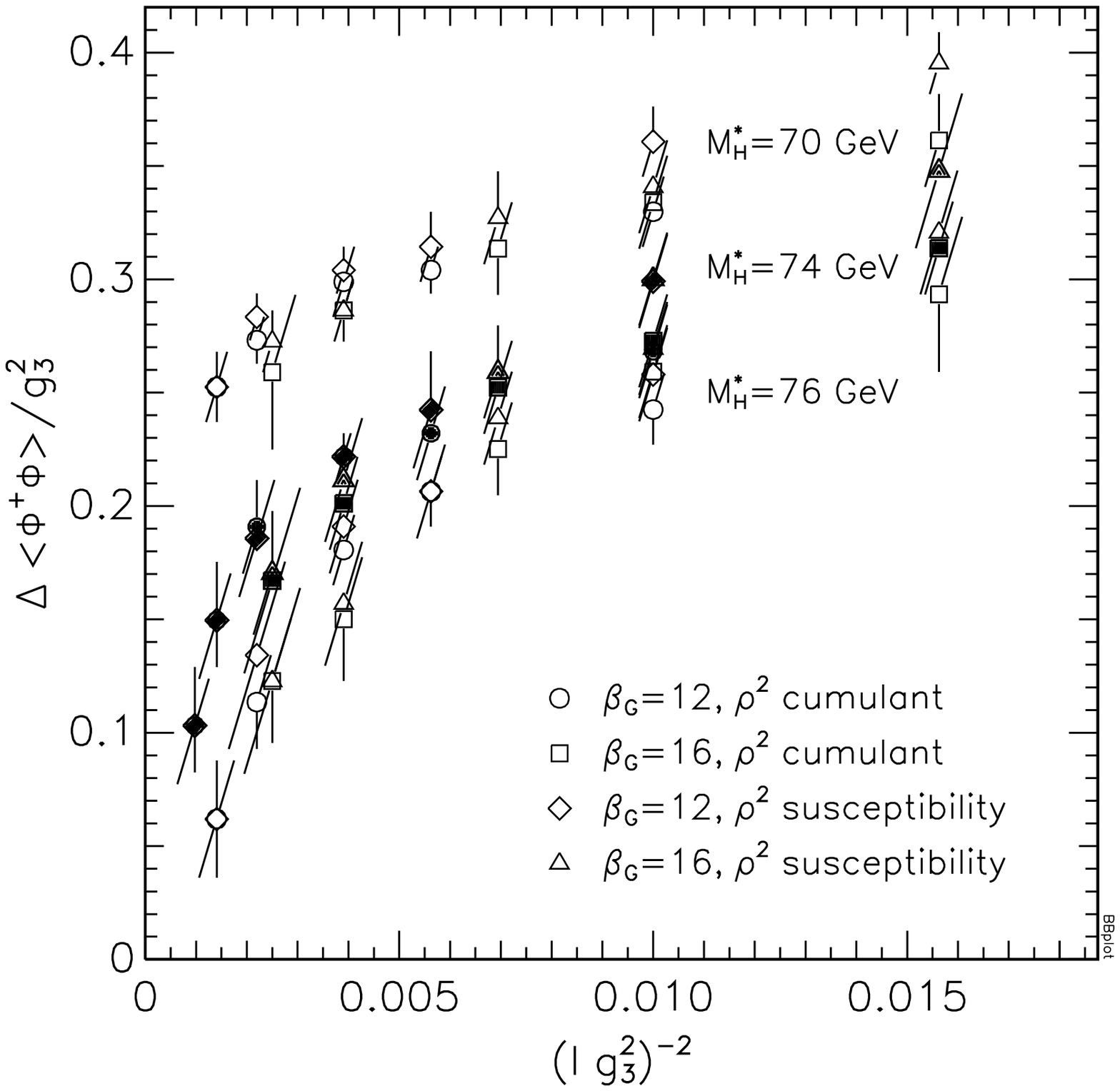,width=5cm,height=5cm,angle=0}
      \vspace{-0.4cm}
      \caption{\sl {\small Thermodynamical limit for $\Delta\langle\phi^+
      \phi\rangle$ at three $M_H^*$ values (here $l=a L$)}}
      \label{fig:lat_mg_fig1.eps}
    \end{center}
  \end{minipage}
  \hfill
  \begin{minipage}[t]{7.3cm}
    \begin{center}
      \epsfig{file=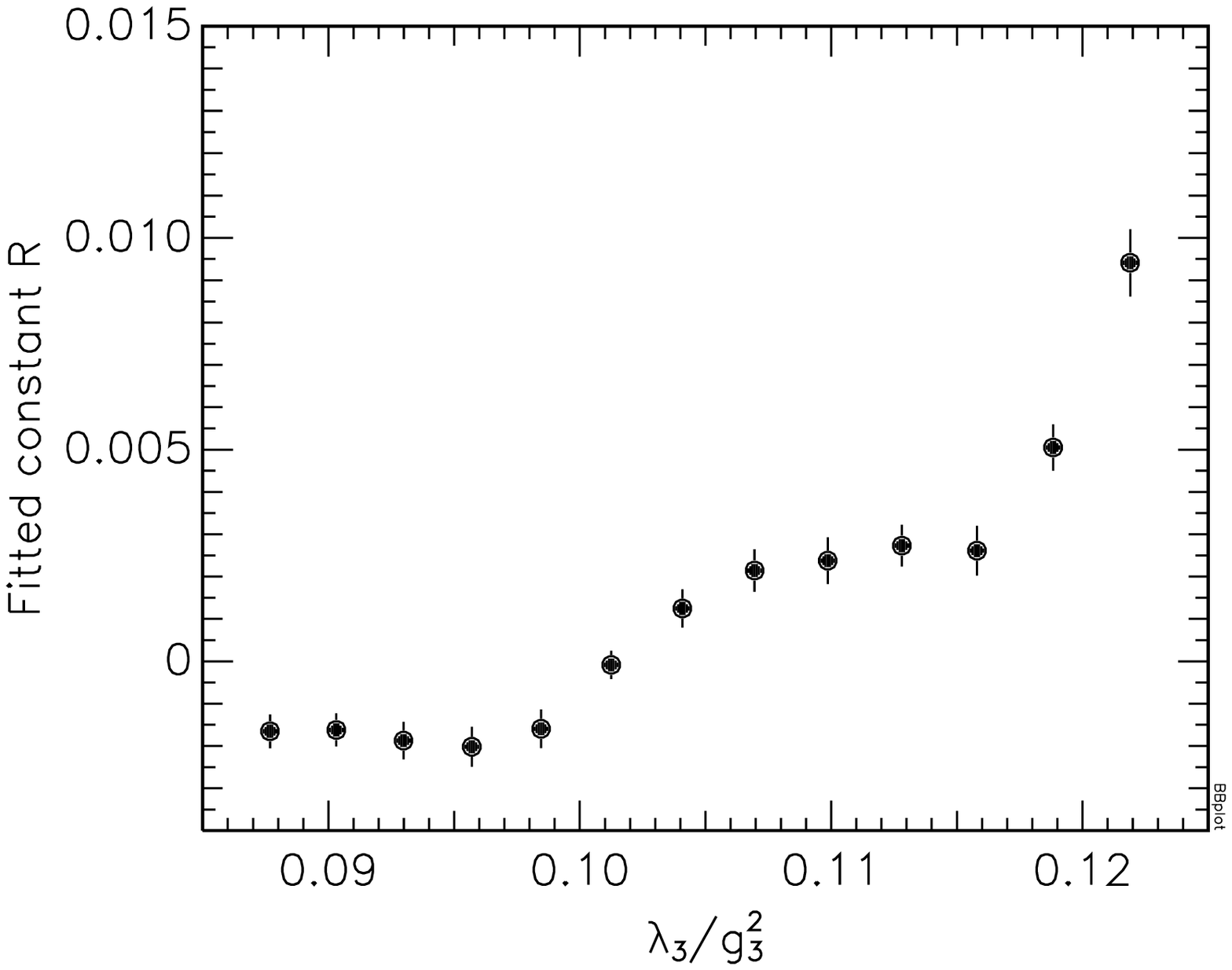,width=5cm,height=5cm,angle=0}
      \vspace{-0.4cm}
      \caption{\sl {\small Minimal distance $R$ of LY zeroes from real axis
      as function of $\lambda_3/g_3^2$}}
      \label{fig:lat_mg_fig5.eps}
    \end{center}
    \end{minipage}
\end{figure}

We have employed
two methods to localise the critical Higgs mass.
First we tried an interpolation of the discontinuity  
\begin{equation}
\Delta \langle \phi^+\phi \rangle/g_3^2 = 1/8\ \beta_G \beta_H 
\Delta \langle \rho^2 \rangle \, ,
\label{eq:jump}
\end{equation}
based on reweighting simulation data taken at $M_H^*=70, 74$
and $76$ GeV, in order to see where this turns to zero. This
discontinuity is directly related to the latent heat.
The pseudocritical $\beta_{H~c}$ has been defined in two ways,
by the maximum of
the $\rho^2$ susceptibility and by the minimum of the
$\rho^2$ Binder cumulant. For each $M_H^*$, an infinite volume extrapolation
must be performed. 
For this purpose we have expressed the {\it physical} lattice size 
by the dimensionless variable $L a g_3^2= 4L/\beta_G$.
We have collected
data from different lattice volumes $L^3$, from 
different $\beta_G$ ($=12$ and $16$) as well as from different
definitions of $\beta_{H~c}$  and found them scattered along
a unique function of this variable (Fig.~\ref{fig:lat_mg_fig1.eps}).
We tried to fit the finite volume 
scaling behaviour as 
$\Delta \langle \phi^+\phi \rangle_{\infty} 
 - \Delta \langle \phi^+\phi \rangle_L \propto (L~a~g_3^2)^{-2}$ 
suggested by the Potts model.
Using only the largest lattice volumes available, the criterion of
vanishing latent heat gives an upper bound 
$\lambda_3^{crit}/g_3^2 < 0.107$ for the existence of a first order 
transition.

The other method detects the end of the transition line 
using the  finite size analysis 
of the Lee--Yang zeroes. 
A genuine phase transition is characterised by {\it non--analytical}
behaviour 
of the infinite volume free energy density. This is caused by zeroes of the 
partition function (in our case as function of $\beta_H$
extended to complex values) 
clustering along a line nearest to 
the real axis and pinching
in the thermodynamical limit. 
If there is a first order phase transition the
first few zeroes are expected at
\begin{eqnarray}
{\mbox{Im}} \beta_H^{(n)} =
 \frac{2 \pi \beta_{H~c}}{L^3 (1+2 \beta_{R~c})
  \Delta \langle \rho^2 \rangle} \left(n-\frac{1}{2}\right), \ \ 
{\mbox{Re}} \beta_H^{(n)} \approx  \beta_{H~c} \ .
\label{eq:pattern}
\end{eqnarray}
The partition function for complex couplings is obtained by reweighting from
measurements at real couplings. The first zeroes can be well localised
using the Newton-Raphson algorithm.
We fit the imaginary part of the first zero for each available length
$L a g_3^2$ according to $ {\mbox{Im}} \beta_H^{(1)}=C (L a g_3^2)^{-\nu}+R$.
The scenario suggested is the change of the first order transition into an
analytic crossover above $M_H^{*crit}$.  A {\it positive} $R$ signals that the
first zero does not approach the real axis in the thermodynamical
limit.  A similar investigation has been performed recently at smaller gauge
coupling in Ref.~\cite{karsch}.
The fitted $R$  crosses zero (Fig.~\ref{fig:lat_mg_fig5.eps}) at
$\lambda_3^{crit}/g_3^2=0.102(2)$.
This corresponds to 
a zero temperature Higgs mass
$m_H^{crit}=72.4(9)$ GeV, and
the phase transition line ends at a
temperature of $T_c=110(1.5)$ GeV. This refers to
the experimental top mass. 
For the purely bosonic Higgs model we get
$m_H^{crit}=67.0(8)$ GeV and 
$T_c=154.8(2.6)$ GeV.
\vspace{-3mm}
\section{Bound states below and above  $m_H^{crit}$}
\vspace{-3mm}
To study 
the ground {\it and} excited bound states
with given $J^{PC}$ 
one has to use cross-correlations between operators ${\cal {O}}_{i}$ forming
a complete set in that channel. 
They describe a spectral decomposition $\Psi_i^{(n)}=\langle \mathrm{vac}|
{\mathcal{O}}_i | {\bf \Psi}^{(n)}\rangle$ with $|{\bf \Psi}^{(n)}\rangle$ 
being 
the (zero momentum) energy eigenstates. According to  
the transfer matrix formalism,
the connected correlation matrix
\begin{eqnarray}
 C_{ij}(t) = \langle {\cal {O}}_{i}(t) {\cal {O}}_{j}(0) \rangle 
 = \sum_{n=1}^{\infty} \Psi_i^{(n)} \Psi_j^{(n)*} e^{-m_n t} 
  \label{eq:spec_dec}
\end{eqnarray}
at time separation $t$ gives the masses {\it and} wave functions.  
 
Practically, only a {\it truncated} set of operators ${\mathcal{O}}_{i}
(i=1,\ldots,N)$ is used assuming that this allows to find the {\it lowest} 
states from the eigenvalue problem  for $C_{ij}(t)$.
However, truncation errors are not small.
Solving instead the generalised eigenvalue problem
\begin{eqnarray}
\sum_j  C_{ij}(t) \Psi_j^{(n)} &=& \lambda^{(n)}(t,t_0) 
\sum_j C_{ij}(t_0) \Psi_j^{(n)} 
  \label{eq:gen_eigen}
\end{eqnarray}
these errors can be kept minimal ($t>t_0, t_0=0,1$)
\cite{luescher}.

The wave function of state $n$ at fixed
small time $t>t_0$ is found to be
$ \Psi_i^{(n)}(t)  = 
\langle \mathrm{vac} | ~{\cal {O}}_i ~e^{-H t} ~| {\bf \Psi}^{(n)} \rangle  $.
The components in the operator basis
characterise the coupling of ${\cal {O}}_i$ 
to the (ground or excited) bound state $n$ in the $J^{PC}$ channel.  
The masses of these states are obtained fitting
the diagonal elements $\mu^{(n)}(t)$
\begin{eqnarray} 
  \mu^{(n)}(t)= \sum_{ij} \Psi_i^{(n)*} C_{ij}(t)
  \Psi_j^{(n)} 
\end{eqnarray}
to a $\cosh$ form with $t$ in some plateau region. 
In contrast to a blocking procedure used in \cite{Philipsen}, 
our base is built by a few types of gauge invariant operators
${\mathcal{O}}_i$, 
properly chosen with respect to lattice 
symmetry and quantum numbers,  
having different  well-defined transverse extensions. 
In the Higgs channel ($0^{++}$) we use the
operators 
$\rho_x^2$ and 
$S_{x,\mu}(l)=\frac{1}{2}{\mathrm{tr}}(\Phi^+_x U_{x,\mu}\ldots 
U_{x+(l-1)\mu,\mu}\Phi_{x+l\mu})$ as well as quadratic Wilson loops of size 
$l \times l$,  in the
W-channel ($1^{--}$) the operators 
$V_{x,\mu}^b(l))=\frac{1}{2}{\mathrm{tr}}(\tau^b\Phi^+_x U_{x,\mu}\ldots
U_{x+(l-1)\mu,\mu}\Phi_{x+l\mu})$
and in the
$2^{++}$ channel $S_{x,\mu}(l)-S_{x,\nu}(l)$
where $l$ expresses the operator extension in lattice units.
In our procedure, the operator $\Phi^{(n)}$ projecting maximally onto 
state 
$n$ is a optimised superposition  
$\Phi^{(n)}=\sum_1^N a_i^{(n)} {\mathcal{O}}_i$ 
with
coefficients provided by the 
(properly normalised) 
solutions $\Psi_i^{(n)}$
of the generalised
eigenvalue equation.
The coefficients give direct access to the spatial extension
of the states under study.
This construction, at different $\beta_G$, should reveal 
the same wave function in position space as function of $l a g_3^2$.
 
\begin{figure}[!htb]
  \begin{minipage}[t]{7.4cm}
    \epsfig{file=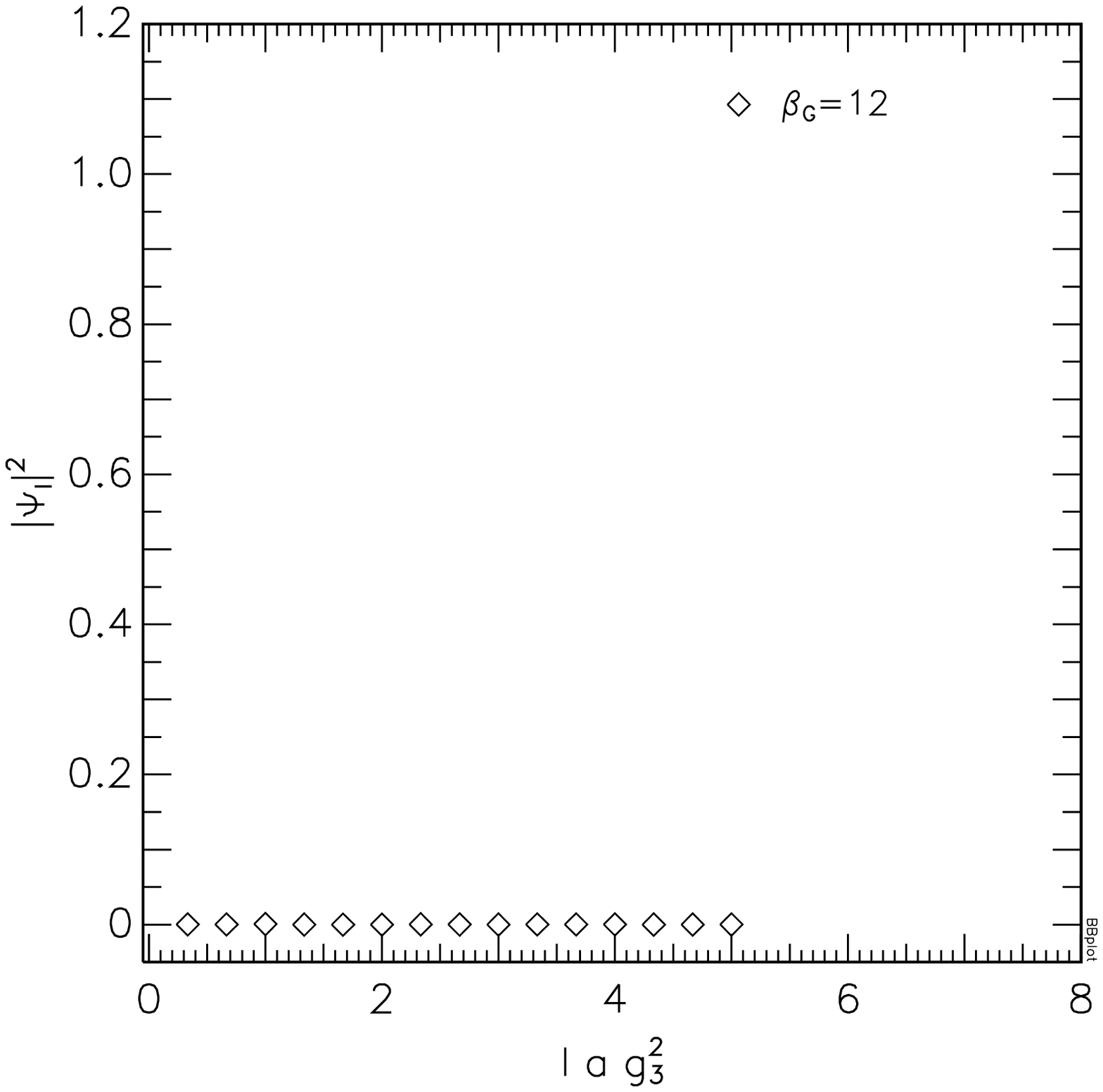,width=3.5cm,height=3.5cm,angle=0}
    \epsfig{file=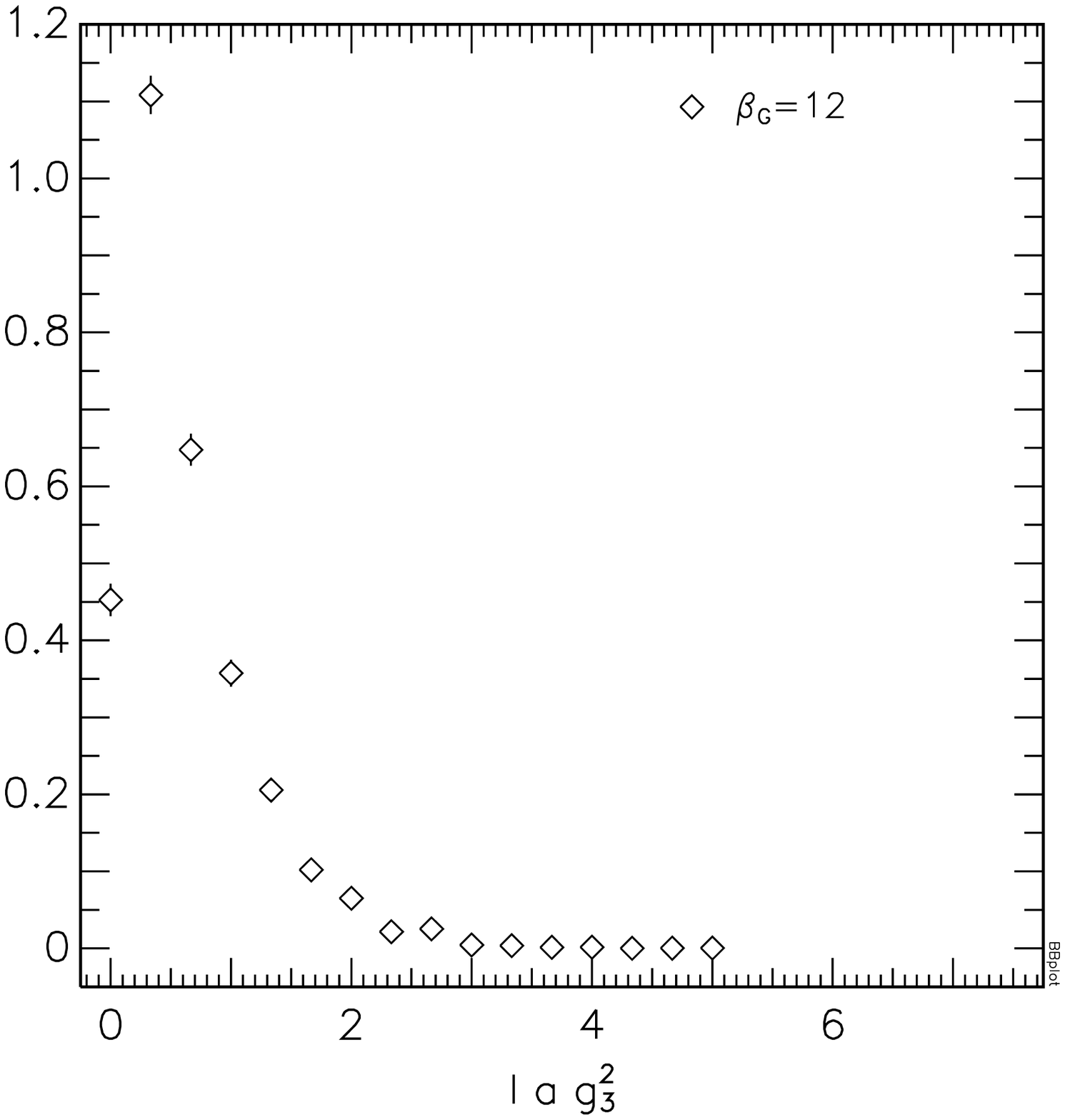,width=3.5cm,height=3.5cm,angle=0}
    \vspace{-5mm}
    \caption{\sl \small{$0^{++}$ ground state, left: Wilson loop projection, 
        right: $S_{x,\mu}(l)$ projection, 
        upper/lower plot: symmetric/Higgs phase}}
    \label{fig:kontwave_10+}
  \end{minipage}
\hfill
  \begin{minipage}[t]{7.4cm}
    \epsfig{file=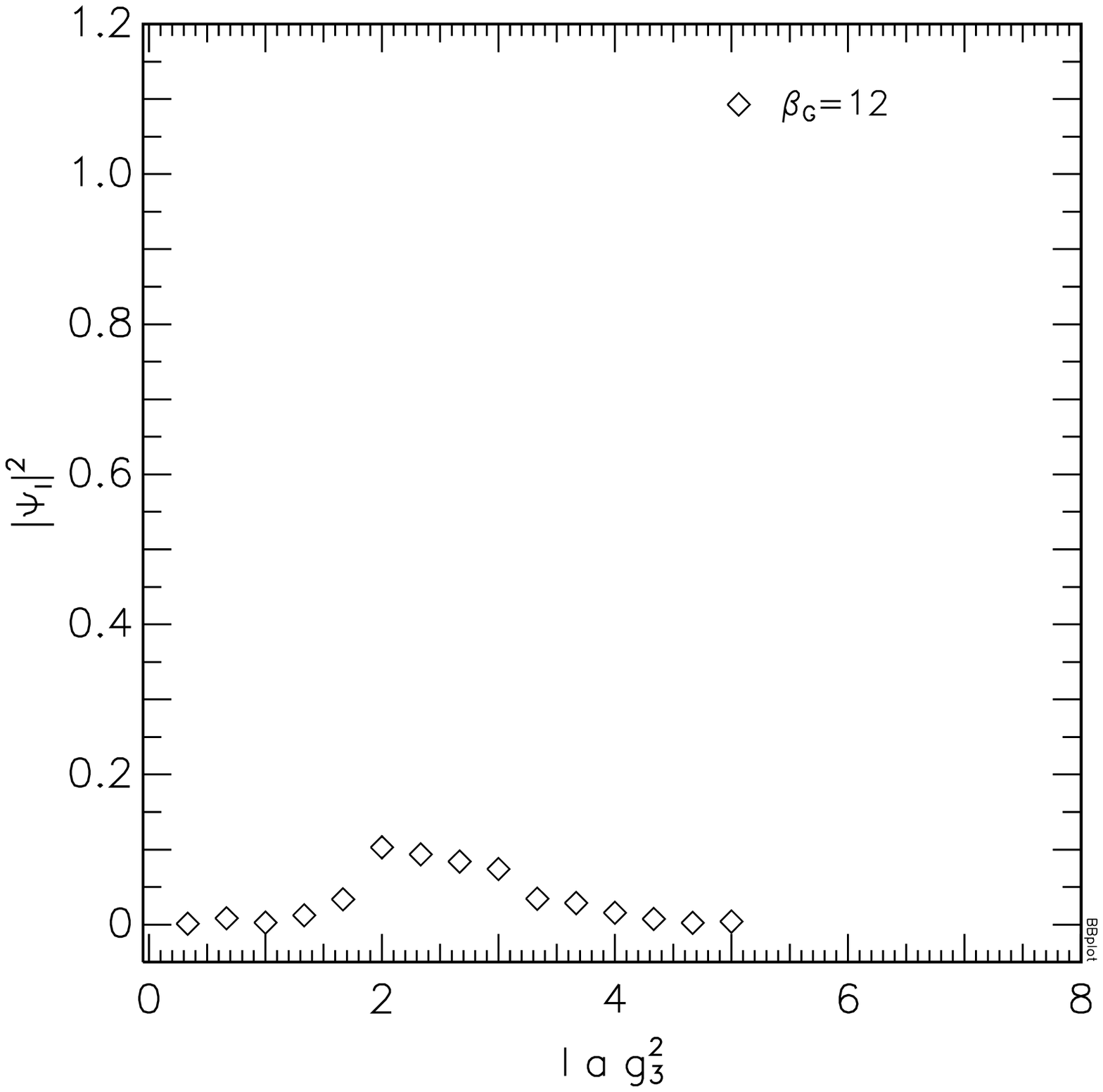,width=3.5cm,height=3.5cm,angle=0}
    \epsfig{file=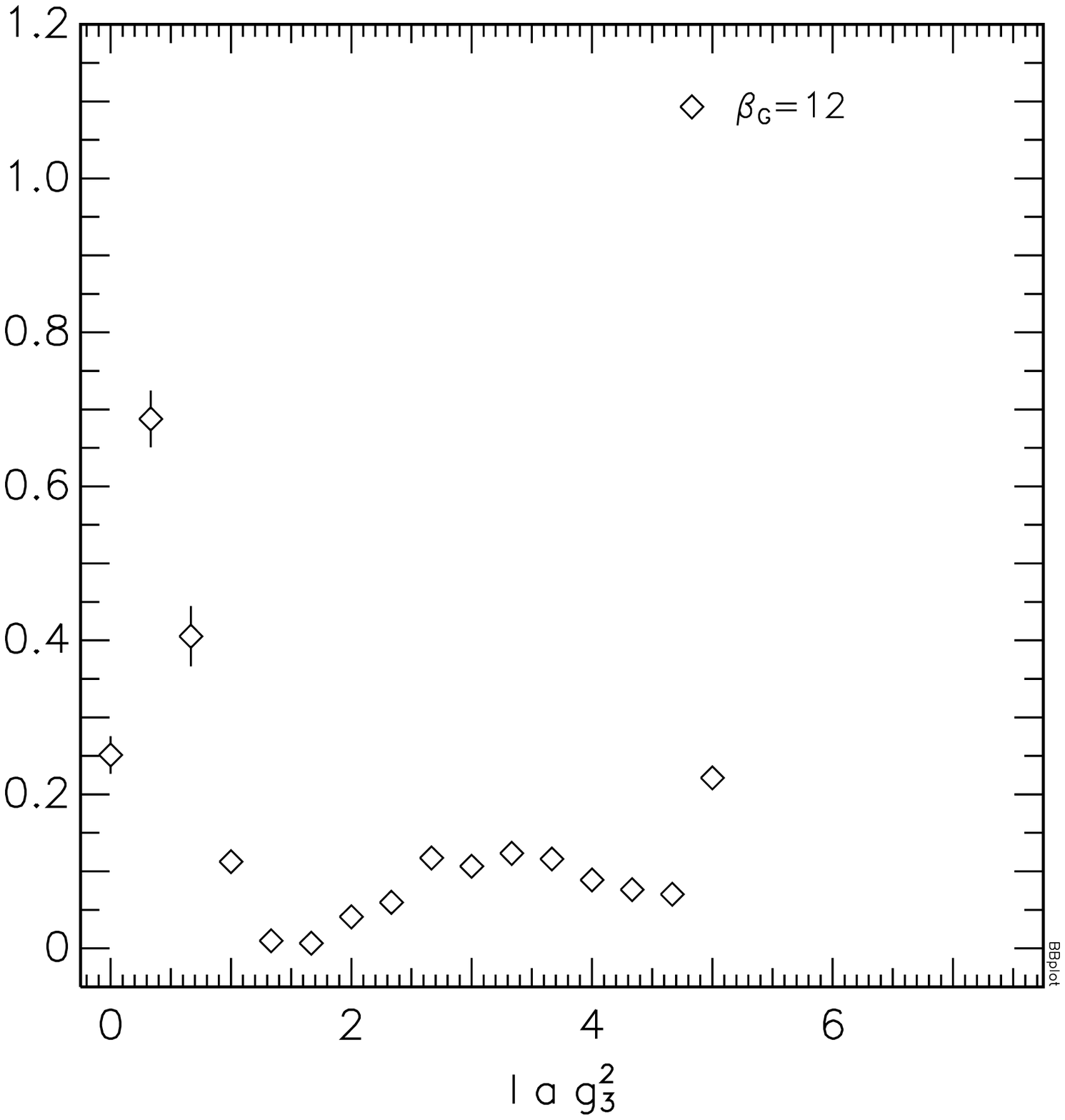,width=3.5cm,height=3.5cm,angle=0}
    \vspace{-5mm}
    \caption{\sl \small{Same as Fig.~3 for second/first excited state}}
    \label{fig:kontwave_30+}
  \end{minipage}
\end{figure}

As examples we present in
Figs.~\ref{fig:kontwave_10+},\ref{fig:kontwave_30+} squared wave functions 
in the $0^{++}$ channel  
vs. $l a g_3^2$ near the end of the phase transition at
$M_H^*=70$ GeV (from a $30^3$ lattice). In
\begin{figure}[!htb]
  \begin{minipage}[t]{7.4cm}
    \epsfig{file=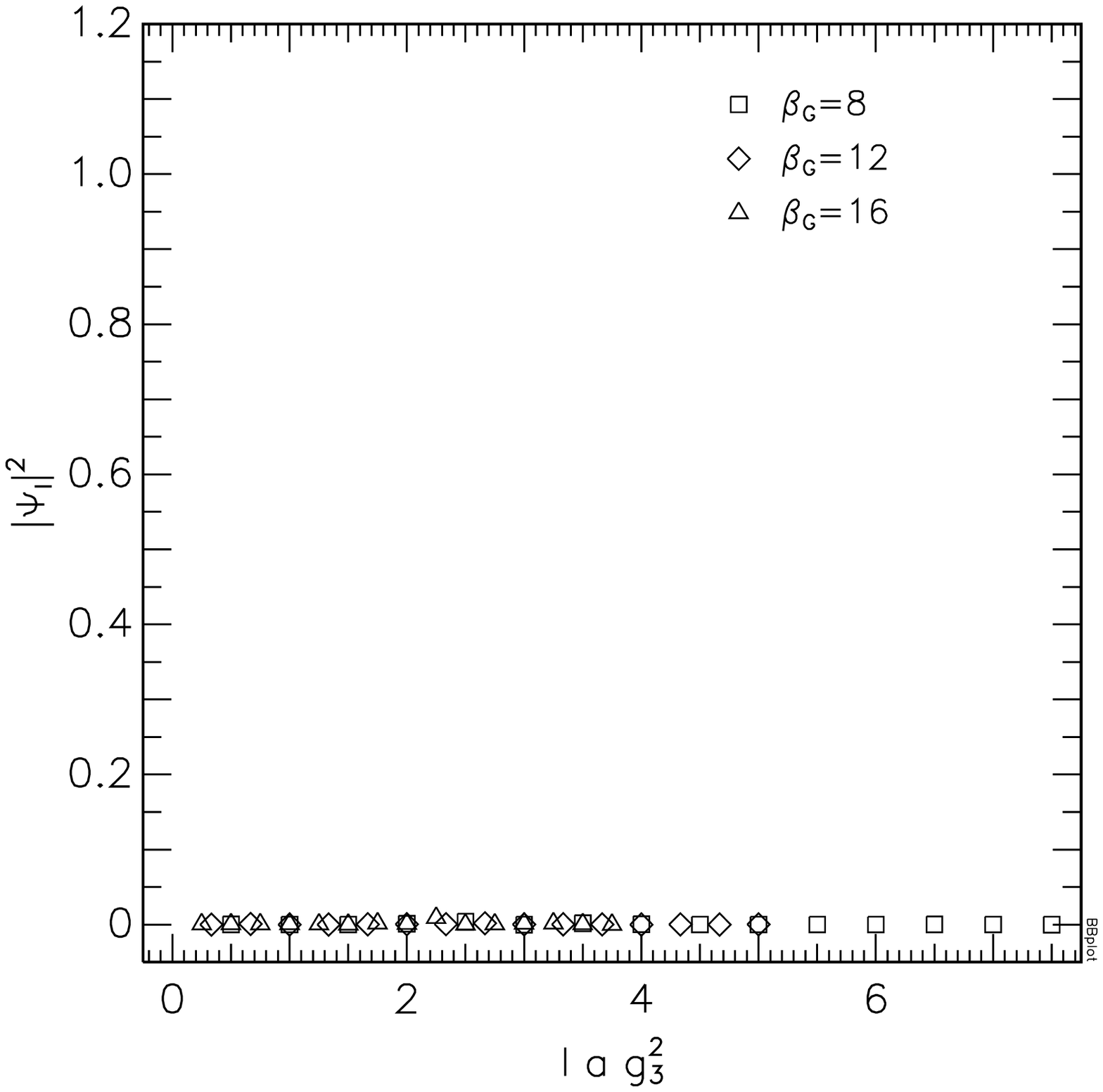,width=3.5cm,height=3.5cm,angle=0}
    \epsfig{file=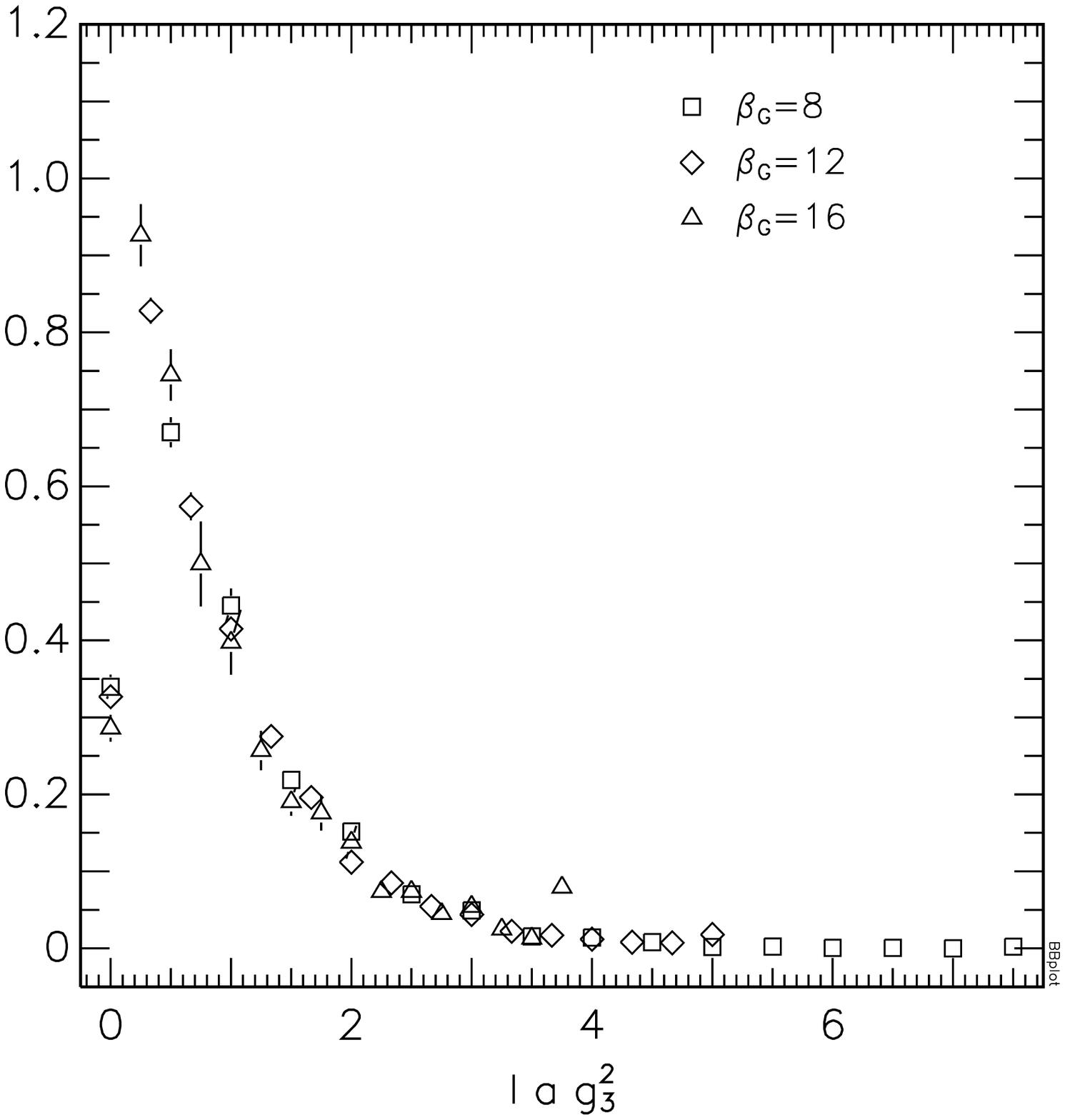,width=3.5cm,height=3.5cm,angle=0}
  \end{minipage}
\hfill
  \begin{minipage}[t]{7.4cm}
    \epsfig{file=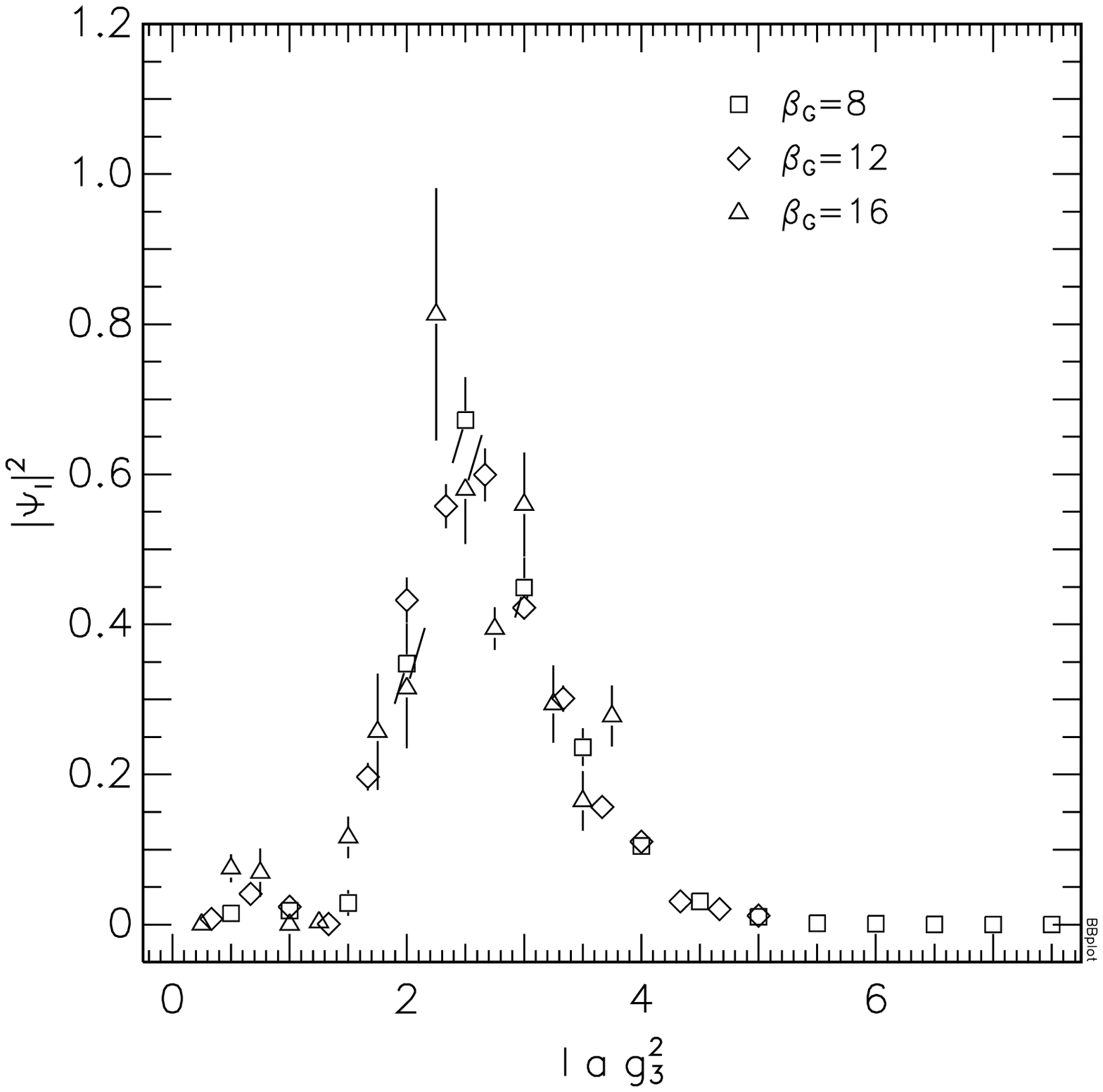,width=3.5cm,height=3.5cm,angle=0}
    \epsfig{file=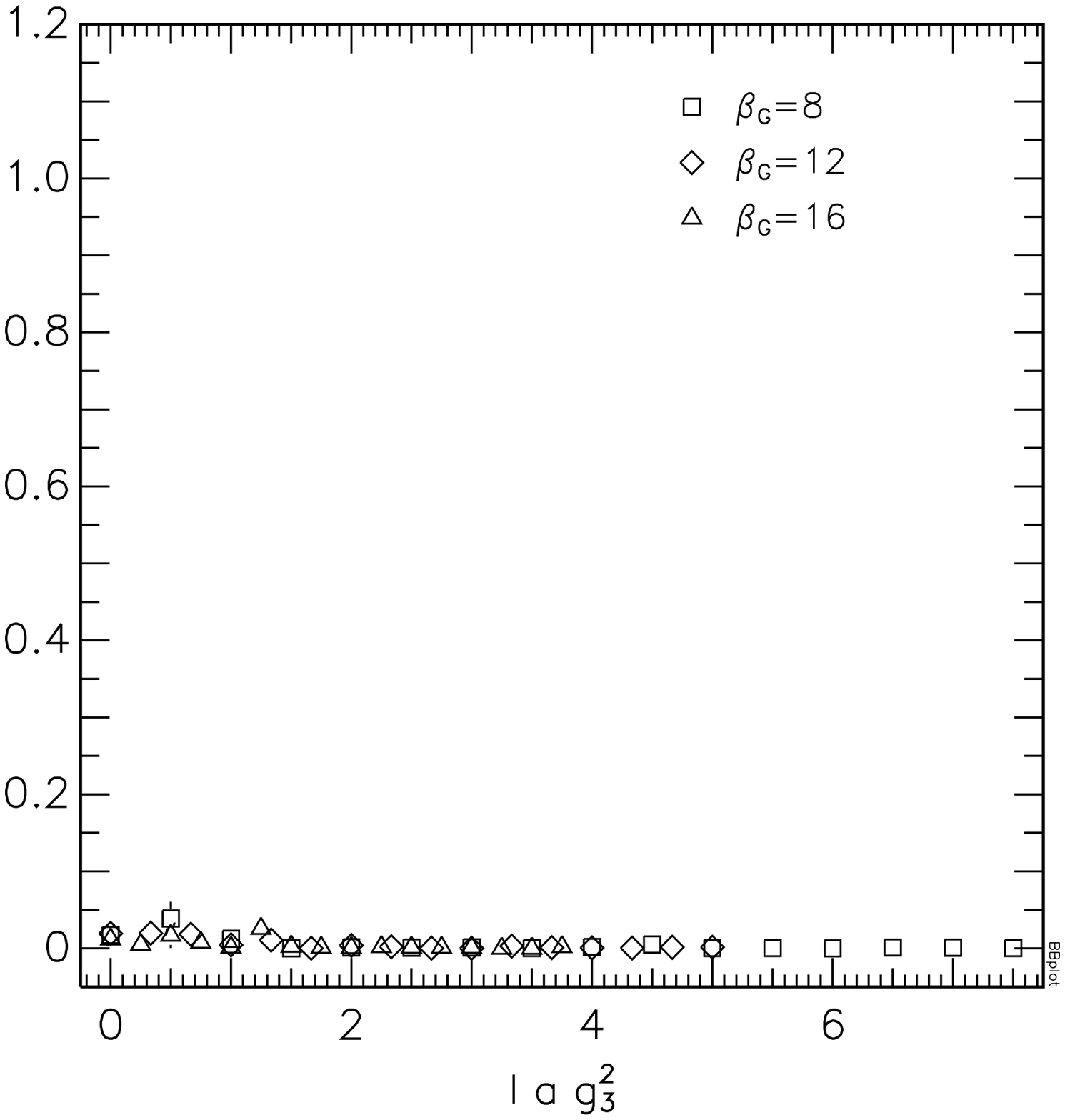,width=3.5cm,height=3.5cm,angle=0}
  \end{minipage}
\end{figure}
\vspace{-8mm}
the symmetric phase the second excitation is a pure $W$-ball state
in agreement with \cite{Philipsen} (the first excitation is a Higgs
state, too). In the broken phase, the first excited Higgs state still
contains some admixture of pure gauge matter
which vanishes only deeply inside this phase.
This seems to be a precursor of the near end of the transition. 
More details, statistics and results for other quantum numbers will be
discussed elsewhere \cite{future}.

Since Wilson loops have different projection properties in both 
phases we have studied the flow of the mass spectrum with
$\beta_H$  or $m_3^2(g_3^2)/g_3^4$  
at $M_H^*=100$ GeV ($\lambda_3/g_3^2=0.15625$) 
safely  
above the critical Higgs mass, too, passing there an analytic crossover. 
 \begin{figure}[!htb]
  \begin{minipage}[t]{14.6cm}
    \begin{center}
      \epsfig{file=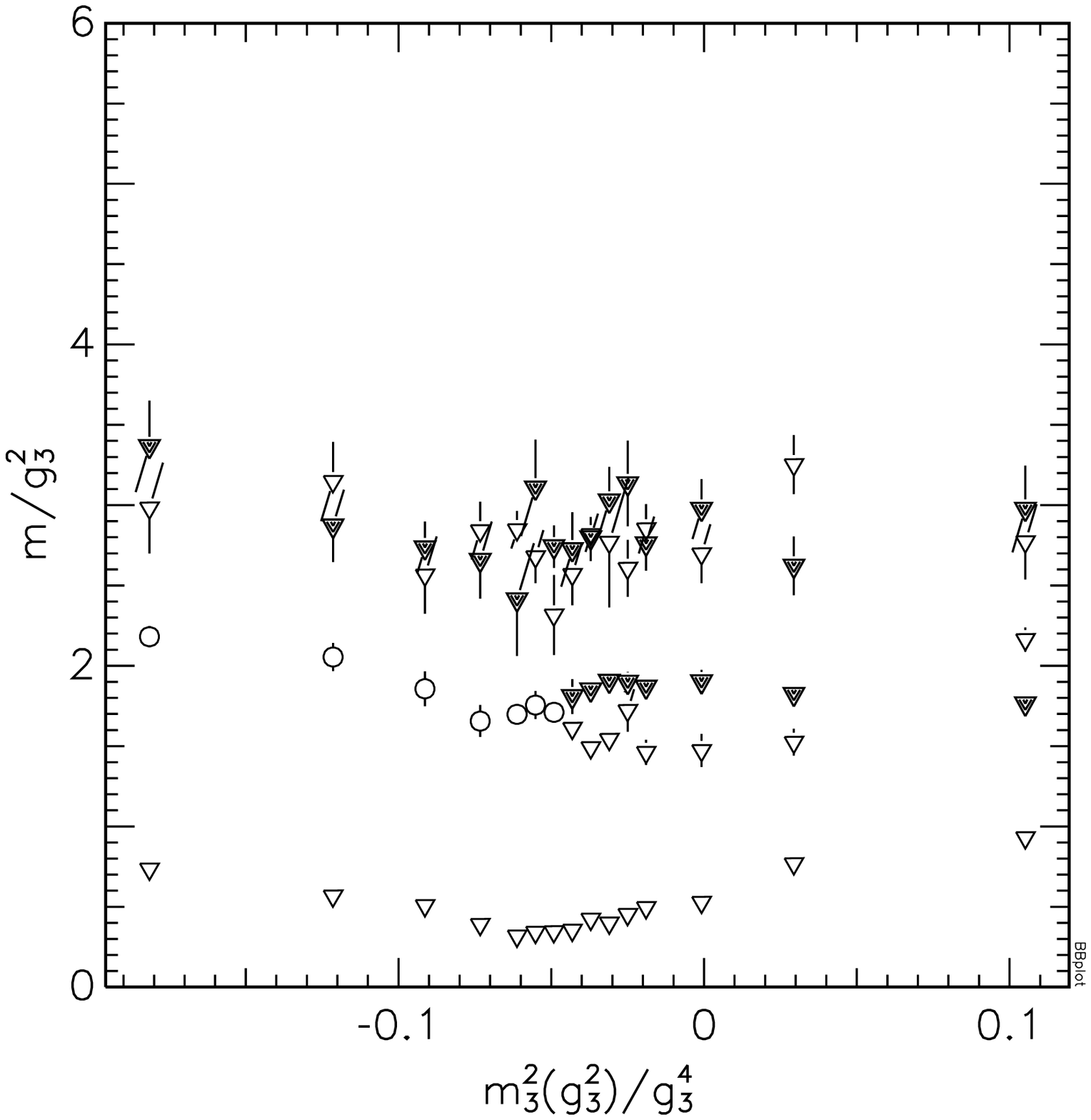,width=5cm,height=5cm,angle=0}
      \epsfig{file=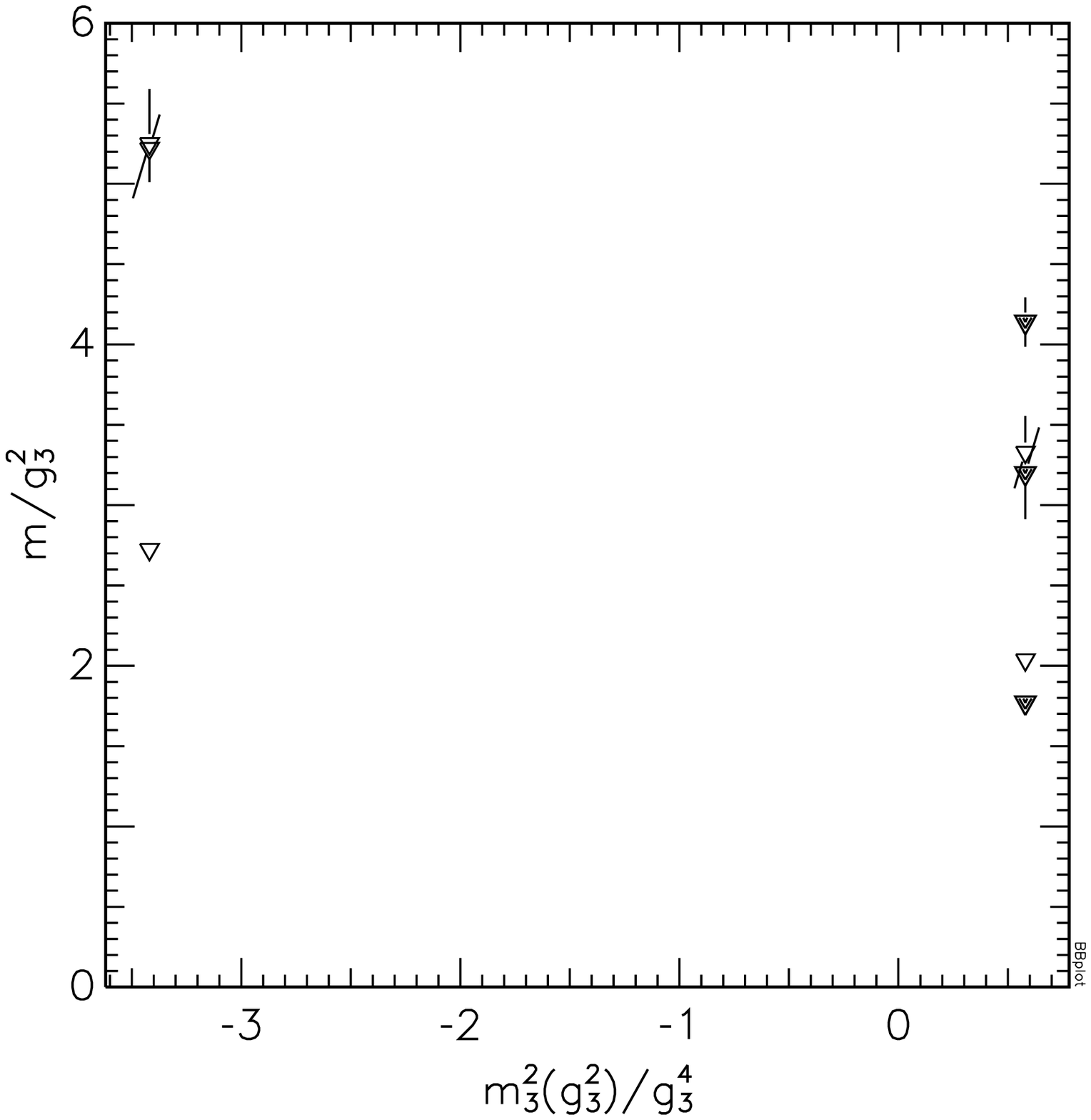,width=5cm,height=5cm,angle=0}
      \vspace{-5mm}
      \caption{\sl{\small $0^{++}$ spectrum beyond the critical Higgs mass;
        left: near to crossover, right: far from crossover. 
        Triangles denote Higgs states, full
        symbols $W$-ball states and circles Higgs states with admixture
        of pure gauge matter}}
      \label{fig:massphi0+}
    \end{center}
  \end{minipage}
\end{figure}
In Fig.~\ref{fig:massphi0+} we present the $0^{++}$ spectrum in the vicinity
and far from the crossover. The wave functions are similar to
those discussed before. Here the spectrum on the ''symmetric'' side 
shows  decoupling of Higgs and $W$-ball
states, too, the lowest $W$-ball mass is roughly independent on $\beta_H$.
Passing the crossover (from ''symmetric'' to the ''broken'' side) the first
Higgs excitation contains also excited glue. These contributions vanish
at lower temperature as indicated by the right 
figure. The minimum of the ground state mass remains finite, therefore 
the correlation length fits into the lattice.

The phenomenological interest is now concentrated on 
extensions of the standard model, with MSSM 
being the most promising candidate. 
Concerning non-perturbative physics in general, however,
the lattice version of the standard Higgs model remains interesting 
as a laboratory for investigating the 
behaviour of hot gauge fields coupled to 
scalar matter, for the characterisation of possible bound states,
for the understanding of real time
topological transitions
and as a cross-check for analytical 
approximation schemes.

\end{document}